\documentclass[a4paper,11pt]{article}
\usepackage{jheppub} 
\usepackage{orcidlink}
\usepackage{mathtools}
\usepackage{amsmath}
\usepackage{amsfonts}
\usepackage{amssymb}
\usepackage{bbold}
\usepackage{bm}
\usepackage{esint}
\usepackage{microtype}

\newcommand{\bn}{{\bf n}}

\newcommand{\bk}{{\bf k}}

\newcommand{\bp}{{\bf p}}

\newcommand{\bq}{{\bf q}}

\newcommand{\GF}{G_{\rm F}}

\newcommand{\exclude}[1]{{}}

\title{Neutrino-antineutrino superfluidity}

\author[a,b]{Damiano F.\ G.\ Fiorillo \orcidlink{0000-0003-4927-9850}}
\affiliation[a]{Istituto Nazionale di Fisica Nucleare (INFN), Sezione di Napoli,
Complesso Universitario di Monte Sant’Angelo, Via Cintia, 80126 Napoli, Italy}
\affiliation[b]{Gran Sasso Science Institute (GSSI), L’Aquila, Italy}

\author[c]{Georg G.\ Raffelt
\orcidlink{0000-0002-0199-9560}}
\affiliation[c]{Max-Planck-Institut f\"ur Physik, Boltzmannstr.~8, 85748 Garching, Germany}

\author[d]{G\"unter Sigl
\orcidlink{0000-0002-4396-645X}}
\affiliation[d]{Universit\"at Hamburg, II.~Institut f\"ur Theoretische Physik, 22761 Hamburg, Germany}

\abstract{Despite their feeble interactions, dense astrophysical neutrinos 
can behave collectively, exchanging flavor through 
waves of the neutrino plasma.
Can collective interactions also induce pairing instabilities and reorganize the neutrino momentum distribution, in analogy to the superfluid instability of fermions? We show that, for standard weak interactions, pairing instabilities can arise only in the presence of discontinuities in the occupation number, such as the sharp Fermi surface responsible for superconductivity in metals. However, discretized energy spectra can mimic such discontinuities and artificially generate superfluid instabilities. These spurious instabilities disappear in the continuum~limit.
}

\begin{document}
\maketitle
\flushbottom

\section{Introduction}

Degenerate fermions with an attractive interaction can form Cooper pairs \cite{Cooper:1956zz}, the mechanism underlying the Bardeen--Cooper--Schrieffer (BCS) theory of superconductivity~\cite{Bardeen:1957mv}. Soon after the emergence of big-bang cosmology and the associated expectation of a cosmological neutrino background, Ginzburg and Zharkov speculated that similar pairings might arise in this astrophysical neutrino environment that could have been highly degenerate~\cite{Ginzburg:1967, Ginzburg:1969}. Since then, the subject has been occasionally revisited from different perspectives \cite{Caldi:1999db, Kapusta:2004gi, Bhatt:2008hr, Addazi:2022kjt, Chodos:2020dgi, Dvornikov:2023nmf}, generally invoking ingredients beyond the Standard Model. The central difficulty is that astrophysical neutrino environments are typically not sufficiently degenerate, and pairing also requires a nonstandard attractive interaction.

In an independent line of development, rooted entirely in Standard-Model interactions, the study of collective neutrino flavor evolution has led to a conceptually related question. In the dense astrophysical environments found in supernovae and neutron-star mergers, neutrinos can exchange flavor collectively through the coherent weak field sourced by the neutrino medium itself \cite{Samuel:1993uw,Samuel:1995ri, Duan:2006an, Sawyer:2004ai, Sawyer:2008zs, Izaguirre:2016gsx, Tamborra:2020cul, Richers:2022zug, Johns:2025mlm, Raffelt:2025wty}. These systems show a variety of collective instabilities \cite{Sawyer:2005jk,Sawyer:2015dsa,Izaguirre:2016gsx,Samuel:1993uw,Duan:2006an,Fiorillo:2024pns,Fiorillo:2025zio,Johns:2021qby}, which can be interpreted as the coherent emission of flavor waves~\cite{Fiorillo:2024bzm, Fiorillo:2024uki, Fiorillo:2024pns, Fiorillo:2025ank, Fiorillo:2025zio, Fiorillo:2025gkw, Fiorillo:2025kko} (see also Ref.~\cite{Johns:2025yxa}). The corresponding quanta, termed flavomons~\cite{Fiorillo:2025npi,Fiorillo:2025gkw,Fiorillo:2026tee}, mediate the flavor exchange by which neutrinos tend to damp differences among lepton number across the energy and angular distribution~\cite{Fiorillo:2024qbl,Fiorillo:2026byh}.

The flavomon field represents the degree of coherence among different flavors in the neutrino gas
\begin{equation}
\Psi_{\alpha\beta}=\sum_\bp\langle a^\dagger_{\beta,\bp} a_{\alpha,\bp}\rangle,    
\end{equation}
where $a_{\alpha,\bp}$ is the annihilation operator of a neutrino with flavor $\alpha$ and momentum $\bp$. The coherent emission of flavomons allows this field to develop a nonvanishing expectation value sourced by the instability. This key insight naturally prompted the question of whether other collective fields could also develop unstable expectation values.

In particular, Refs.~\cite{Volpe:2013uxl,Serreau:2014cfa} considered the neutrino–antineutrino pairing correlator
\begin{equation}
    \Delta_{\alpha\beta}=\sum_\bp \langle a_{\alpha,\bp}b_{\beta,-\bp}\rangle,
\end{equation}
where $b_{\alpha,\bp}$ is the annihilation operator of an antineutrino with momentum $\bp$ and flavor $\alpha$. This is precisely the superfluidity correlator for neutrinos and antineutrinos. Serreau and Volpe~\cite{Serreau:2014cfa} emphasized a particularly interesting point: this neutrino-antineutrino pairing function is directly sourced by the neutrino and antineutrino densities. By contrast, the neutrino-neutrino pairing function $\langle a_{\alpha,\bp} a_{\beta,-\bp}\rangle$, corresponding to standard BCS theory, can only become unstable through an initial seed. Nevertheless, they did not explicitly determine whether this sourcing mechanism was sufficient for the pairing function to grow to a macroscopically significant level. Ref.~\cite{Kartavtsev:2015eva} later concluded that this would not happen due to the large scale separation between the oscillation frequency of the pairing correlators, determined by the kinetic neutrino energy, and the driving rate of the correlators, determined by the potential energy of self-interaction. Indeed, a harmonic oscillator driven at a rate much slower than its natural frequency can only reach amplitudes suppressed by the ratio between these two rates.

Very recently, this question was revisited by Huang and Wu~\cite{Huang:2026fcb}, who raised the interesting question of whether an instability capable of driving the growth of $\Delta_{\alpha \beta}$ can arise in practice. Such a solution would circumvent the argument of Ref.~\cite{Kartavtsev:2015eva} by allowing for exponential growth in the pairing function. Huang and Wu's analysis showed that, within a discretized representation of momentum space, such instabilities appear quite generically.

Here, we examine this question from a different perspective, namely through its connection to the original BCS picture of fermionic superfluidity. In agreement with Cooper's original results~\cite{Cooper:1956zz}, we show that an instability induced by a very feeble attractive interaction requires a sharp Fermi surface. The discontinuity in the occupation number then provides a reservoir of states that can pair at essentially no energy cost, since they all have the same energy. This conclusion does not contradict the findings of Ref.~\cite{Huang:2026fcb}, because a set of discrete energy levels effectively mimics multiple Fermi surfaces, each characterized by a discontinuity in the occupation number.

We show through an explicit example that, for ever finer
discretization, the instability disappears. Overall, our results suggest that neutrino-antineutrino pairing, even if it is sourced in dense environments as found in Refs.~\cite{Serreau:2014cfa, Huang:2026fcb}, cannot develop instabilities unless neutrinos are highly degenerate with a temperature $T\lesssim \sqrt{2} \GF n_\nu$, where $\GF$ is Fermi's constant and $n_\nu$ is the neutrino number density.

\section{General expectations from the theory of superconductivity}\label{sec:superconductivity}

Our starting point for connecting to the BCS theory of superconductivity is the exact many-body neutrino Hamiltonian density
\begin{equation}\label{eq:Hamiltonian}
    \mathcal{H}_{\rm int}=\frac{\GF}{\sqrt{2}}\sum_{\alpha,\beta}\overline{\nu}_\alpha\gamma^\mu P_L \nu_\alpha \overline{\nu}_\beta \gamma_\mu P_L \nu_\beta,
\end{equation}
where $\alpha$ denotes flavor, $P_L=(1-\gamma^5)/2$ is the projector on the left-handed spinor, and we assume neutrinos to be massless Weyl spinors. This Hamiltonian describes an interaction mediated by a vector field between neutrinos and antineutrinos. Therefore, in principle, it might induce pairing (i)~among equal species, i.e., neutrino--neutrino pairing, described by a structure function $\langle a_{\alpha,\bp} a_{\beta,-\bp}\rangle$, and (ii)~among different species, i.e., neutrino--antineutrino pairing, described by a structure function $\langle a_{\alpha,\bp} b_{\beta,-\bp}\rangle$. 

While the early works on cosmological neutrino superfluidity have focused on the former~\cite{Ginzburg:1967, Ginzburg:1969}, the more recent works in the context of flavor conversion of supernova neutrinos have instead focused on the latter~\cite{Serreau:2014cfa,Huang:2026fcb}. There is a general reason for this divergence: the exchange of flavor is mediated by a nonvanishing $\langle \overline{\nu}_\alpha(x) \gamma_\mu \nu_\beta(x)\rangle$. Such a nonvanishing expectation value can emerge from an average flavor field $\langle a^\dagger_{\beta,\bp} a_{\alpha,\bp}\rangle$ in the neutrino sector, equivalent to \smash{$\langle b^\dagger_{\alpha,\bp}a_{\beta,\bp}\rangle$}, what would be called a particle--hole condensate in many-body physics. Or it can emerge from an average neutrino--antineutrino pairing \smash{$\langle a_{\alpha,\bp} b_{\beta,-\bp}\rangle$}, namely a particle--antiparticle condensate. This is the main possibility considered by Serreau and Volpe~\cite{Serreau:2014cfa}; since the current $\langle \overline{\nu}_{\alpha}(x) \gamma_\mu \nu_\beta(x)\rangle$ is already nonzero, they showed that it also sources the particle--antiparticle condensate.

However, while the particle--particle condensate $\langle a_{\alpha,\bp} a_{\beta,-\bp}\rangle$ is not sourced by the neutrino and antineutrino densities, it may equally well develop an instability. The lack of a direct source alone is not sufficient reason to dismiss this possibility. Overall, therefore, one should really examine four different kinds of coherent effects:
\begin{itemize}
    \item the particle--hole $\langle a^\dagger_{\beta,\bp}a_{\alpha,\bp}\rangle $ (or antiparticle--antihole $\langle b^\dagger_{\alpha,\bp}b_{\beta,\bp}\rangle$) condensate, corresponding to the flavomon field that causes collective flavor exchange. (More specifically, only the $\alpha\neq \beta$ elements correspond to flavomons. The $\alpha=\beta$ elements are the neutrino densities of different flavors; oscillations in the traceless part are what we called neutrino plasmons~\cite{Fiorillo:2025npi}, while coherent oscillations in the total density $\sum_\alpha \langle a^\dagger_{\alpha,\bp} a_{\alpha,\bp}\rangle$ represent the standard plasmon.) The traditional interpretation of this field is similar to that of spin waves, where spin oscillates coherently in space and time. It is now widely understood that in several circumstances, flavomons can be emitted in a stimulated fashion by neutrinos, causing the usual flavor instabilities;
    \item the particle--antihole $\langle a^\dagger_{\beta,\bp}b_{\alpha,\bp}\rangle$ condensate describes, in principle, coherent oscillations of neutrinos into antineutrinos. Sawyer~\cite{Sawyer:2022ugt, Sawyer:2023dov} suggested that this condensate might also become unstable, leading to collective neutrino--antineutrino conversions. However, as we showed earlier~\cite{Fiorillo:2024wej}, such an instability is prevented by helicity conservation that disables the coherent conversion $\nu_\bp+\overline{\nu}_{-\bp}\to \nu_{-\bp}+\overline{\nu}_\bp$. Such an instability can only occur when neutrino mass is included, causing an additional $m_\nu/|\bp|\ll 1$ suppression;
    \item the particle--antiparticle $\langle a_{\alpha,\bp}b_{\beta,-\bp}\rangle$ condensate, corresponding to superfluid neutrino--antineutrino pairing. Since neutrino--antineutrino interaction in the Standard Model is attractive, this possibility is conceptually open. Serreau and Volpe~\cite{Serreau:2014cfa} showed that the conventional neutrino and antineutrino densities actually source this condensate. If an instability exists, as speculated in Ref.~\cite{Huang:2026fcb}, it thus would have natural seeds. Whether such an instability indeed exists is the main subject of our work;
    \item the particle--particle $\langle a_{\alpha,\bp}a_{\beta,-\bp}\rangle$ (and antiparticle--antiparticle $\langle b_{\alpha,\bp}b_{\beta,-\bp}\rangle$) condensate, corresponding to superfluid neutrino and antineutrino pairing. Notice that, in analogy to Serreau and Volpe~\cite{Serreau:2014cfa}, the particle--antihole condensate $\langle a^\dagger_{\beta,\bp} b_{\alpha,\bp}\rangle$ might seed the particle--particle condensate, although, as we have seen, we do not expect the particle--antihole condensate to exist in the first place. Based on standard BCS theory, we expect that the particle--particle condensate might in principle become unstable if an attractive neutrino--neutrino and antineutrino--antineutrino interaction were to exist, which is not the case in the Standard Model. Moreover, our arguments against the neutrino--antineutrino pairing condensate apply equally to the particle--particle one, suggesting that this instability cannot exist in realistic environments.
\end{itemize}

A mostly formal clarification is needed on the definition of pairing functions. Since the operators $a_{\alpha,\bp}b_{\beta,-\bp}$ do not conserve the total number of particles, rigorously a pairing function can never develop in a state with a fixed particle number. However, in the thermodynamic limit of a large number of particles, there is no practical change if one replaces a state with a fixed particle number by e.g.\ a coherent state with a slight spread in the particle number. This issue emerges identically in the theory of superfluidity of the Bose liquid, or in the superconductivity of the charged Fermi liquid. One may also dispense from introducing particle number violation, and formulate the theory in a canonical ensemble with a fixed particle number. In that case, the pairing correlator is defined as the coefficient of the pole of the two-particle Green's function. This more formal description of superconductivity (see, e.g., Ref.~\cite{bogoliubov1970lectures}) leads to exactly the same results as the much simpler physical treatment that we refer to here.

With this premise, we turn to the question of when a particle--antiparticle condensate might develop. The basic notion of BCS theory is that an attractive interaction can lead to an instability, whereby the pairing function $\langle a_{\alpha,\bp}b_{\beta,-\bp}\rangle$ becomes nonzero. The nature of the instability is simply the formation of a bound state mediated by the attractive interaction. This is most simply elucidated by Cooper's argument~\cite{Cooper:1956zz} that two particles introduced in an otherwise filled metal can form a bound state. 

The most striking feature of superconductivity is that an arbitrarily weak interaction suffices to produce such a bound state. In quantum mechanics, a particle with a kinetic energy $K$, immersed in an attractive potential with strength $U\ll K$ (e.g.~a spherical potential well of depth $U$) does not develop bound states. It takes a depth comparable with $K$ to confine the particle. For example, within a spherical well of spatial size $a$, a nonrelativistic particle of mass $m$ has a minimum kinetic energy $K\sim 1/(m a^2)$. Therefore, unless $U\gtrsim 1/(m a^2)$, there will be no bound states. An exception to this rule is found in systems of lower dimensionality. For example, the textbook problem of a one-dimensional potential well possesses bound states for arbitrarily weak attractive potential. However, a metal is a three-dimensional system, raising the question of how can a very weak interaction induce pairing of electrons in a superconductor.

The answer was provided by Cooper's argument~\cite{Cooper:1956zz} that it is really a sharp discontinuity in the density of states that allows particles to pair even when the interaction is arbitrarily weak. For example, in a metal, the electron phase-space distribution has a sharp discontinuity at the Fermi surface $p^2/2 m=\mu$, where $m$ is the electron mass, $p$ its momentum, and $\mu$ its chemical potential. Particles exactly at the Fermi surface have all the same energy. Therefore, while their kinetic energy $K\sim \mu$ may far exceed their mutual attractive potential~$U$, it takes an infinitesimal energy to restructure their wavefunction on the Fermi surface, where all have the same kinetic energy. If two new electrons are added to the system, one can construct a bound state between them, using a superposition of correlated plane waves lying on the same Fermi surface; because they have the same energy, an infinitesimal $U$ is sufficient for this purpose.

The general argument that a discontinuity in the density of states is what facilitates bound states also explains why a potential well in one dimension can achieve this. Since a nonrelativistic particle has $K=p^2/2m$, the density of states (the number of states per unit length $L$) is
\begin{equation}
    \frac{dN}{dL}=\frac{dp_z}{2\pi}=\frac{dK}{2\pi}\sqrt{\frac{2m}{K}}.
\end{equation}
There is an infinite density of states at arbitrarily small $K$, which can always be reshaped into a bound state by an arbitrarily weak $U$. This is to be contrasted with three dimensions, where the density of states (the number of states per unit volume~$V$) is
\begin{equation}
    \frac{dN}{dV}=\frac{d^3\bp}{(2\pi)^3}=dK \frac{(2m)^{3/2}\sqrt{K}}{2\pi^2},
\end{equation}
so there is a vanishing number of states with zero kinetic energy.

Based on these arguments, we generally expect that the attractive neutrino--antineutrino interaction, with a typical potential energy $U\sim \sqrt{2}\GF n_\nu$, can produce bound states only if it is comparable with the typical neutrino kinetic energy $K\sim \mu_\nu$, where $\mu_\nu$ is the neutrino chemical potential, or in a nondegenerate system $K\sim T$, where $T$ is the temperature. On the other hand, if neutrinos are strongly degenerate, then they offer a Fermi surface smeared over a range $\delta K\sim T$; in this case, an attractive interaction with $U\sim T$ may suffice to establish a pairing instability and the formation of bound pairs. We now turn to more formal arguments supporting this view.

\section{Mean-field evolution equations}

With the goal of deriving the mean-field equations for our system, we return to the Hamiltonian of Eq.~\eqref{eq:Hamiltonian}. Schematically it takes the form, already reduced to normal ordering,
\begin{eqnarray}\label{eq:Hamiltonian-1}
    \mathcal{H}_{\rm int}&=&\frac{\GF}{\sqrt{2}}\sum_{1,2,3,4}T_{12;34}\left[-a^\dagger_1 a^\dagger_3a_2 a_4+2a^\dagger_1 b^\dagger_4 a_2 b_3+2a^\dagger_1 b^\dagger_2b_3 a_4+2a^\dagger_1 a_2 b_3 a_4\right. \nonumber\\ &&\left.\kern0em{}+2a^\dagger_1 a^\dagger_3 b^\dagger_4 a_2 -2b^\dagger_2 b_1 b_3 a_4-2b^\dagger_2 a^\dagger_3 b^\dagger_4b_1+b_1 a_2 b_3 a_4+a^\dagger_1 b^\dagger_2 a^\dagger_3 b^\dagger_4-b^\dagger_2 b^\dagger_4 b_1 b_3\right],
\end{eqnarray}
where we denote by $i=1, 2,3,4$ a collective index that marks both the momentum $\bp_i$ and the flavor state $\alpha_i$ of the neutrino. The spinor matrix element is
\begin{equation}
    T_{12,34}=\overline{u}_{\bp_1}\gamma_\mu u_{\bp_2}\overline{u}_{\bp_3}\gamma^\mu u_{\bp_4},
\end{equation}
where for massless neutrinos, the spinor for a neutrino with momentum $\bp$ is identical to the one for an antineutrino with the same momentum and opposite helicity, as we have emphasized previously~\cite{Fiorillo:2024wej}. The sum is restricted to momenta satisfying conservation, for example $\bp_1+\bp_3=\bp_2+\bp_4$ for the term $a^\dagger_1 a^\dagger_3 a_2 a_4$, or $\bp_1+\bp_2+\bp_3+\bp_4=0$ for $b_1 a_2 b_3 a_4$. 

In the mean-field approximation, we assume that the particles attain the nonvanishing expectation values $\langle a^\dagger_2 a_1\rangle=\rho_{12}$, $\langle b^\dagger_1 b_2\rangle = \overline\rho_{12}$, and $\langle a_1 b_2\rangle=\Delta_{12}$. Notice that for neutrinos we have used the transposed convention compared to the usual one. Furthermore, nothing prevents in general the possibility of having also nonvanishing expectation values $\langle b^\dagger_1 a_2\rangle$, $\langle a_1 a_2\rangle$, and $\langle b_1 b_2\rangle$. The first correlation function $\langle b^\dagger_1 a_2\rangle$ is never sourced at the refractive level due to helicity conservation~\cite{Fiorillo:2024wej}. The pairing functions $\langle a_1 a_2 \rangle$ and $\langle b_1 b_2 \rangle$ are not sourced due to the same argument that we present below for $\langle a_1 b_2\rangle$.

The mean-field approach means that fermions are assumed to move in the mean field provided by all the others. On the level of the 
Hamiltonian of Eq.~\eqref{eq:Hamiltonian-1}, this requires one to take the expectation value of different operator pairs, bringing it to the general form
\begin{equation}
    \mathcal{H}_{\rm int}=h_{\alpha \beta} a^\dagger_\alpha a_\beta+\overline h_{\alpha \beta} b^\dagger_\alpha b_\beta + \Phi^*_{\alpha \beta} a_\alpha b_\beta+\Phi_{\alpha \beta}b^\dagger_\beta a^\dagger_\alpha.
\end{equation}
The explicit form of the mean-field matrices $h$, $\overline h$ and $\Phi$ follows from the structure of the Hamiltonian.

To examine the possibility of an instability, we first notice that it cannot couple, at linear level, the pairing function $\Delta$ with its conjugate $\Delta^\dagger$ or the density $\rho$ and $\overline\rho$. In the absence of interactions, the components of $\Delta_{12}$ oscillate with frequency $E_{\bp_1}+E_{\bp_2}$, where $E_\bp=|\bp|$ is the neutrino energy, which is widely separated from the oscillation frequency of $\Delta^\dagger$, $\rho$, and $\overline{\rho}$. Therefore, at linear level, it is enough to consider the behavior of $\Delta_{12}$ alone. We stress that the other terms, which do not contain $\Delta_{12}$, need not vanish; they simply oscillate at a completely different frequency and therefore cannot participate in the instability. From the mean-field Hamiltonian, we immediately deduce the equation for its time evolution in matrix form
\begin{equation}
    i\frac{d\Delta}{dt}=(E_{\bp_1}+E_{\bp_2})\Delta+h \Delta+\Delta \overline{h}^T+\Phi-\Phi \overline\rho-\rho \Phi.
\end{equation}

To proceed, we need the explicit form of $h$, $\overline h$, and $\Phi$. This task simplifies considerably at the linear level in $\Delta$, which is all we need to find an instability. Since the terms proportional to $h$ and $\overline{h}$ already contain one power of $\Delta$, they can be computed from the densities $\rho$ and~$\overline\rho$ with $\Delta=0$. Thus we find
\begin{equation}
    h_{12}=-\overline{h}_{12}=\sqrt{2}\GF \sum_{1,2,3,4}(\rho_{43}-\overline{\rho}_{43})(T_{12,34}-T_{14,32}),
\end{equation}
where in these remaining terms, the summation over momenta is constrained only by the condition $\bp_1+\bp_3=\bp_2+\bp_4$, which is momentum conservation in the particle--hole channel. These equations encode the usual refractive energy shifts of neutrinos and antineutrinos, which engender their collective flavor evolution. For our purposes, their explicit form is not particularly relevant. In a single-flavor context, they simply represent a small renormalization of the neutrino and antineutrino energies by opposite amounts, of the order of $\GF (n_\nu-n_{\overline{\nu}})$, negligible compared with the much larger kinetic energies $E_{\bp_1}+E_{\bp_2}$.

The function $\Phi$ multiplies the unperturbed densities $\rho$ and $\overline{\rho}$, so in $\Phi$, we must keep terms of order $\Delta$. On the other hand, the terms proportional to $\rho$ and $\Delta^\dagger$ can be neglected, since they oscillate at vastly different frequencies, as mentioned earlier, and therefore cannot participate in a possible unstable mode. If we collect in the definition of $\Phi_{12}$ only the terms proportional to $\Delta$, we find
\begin{equation}
    \Phi_{12}=\sqrt{2}\GF\sum_{1,2,3,4}\Delta_{43}(T_{12,34}-T_{14,32}),
\end{equation}
where the condition of momentum conservation is $\bp_1+\bp_2=\bp_3+\bp_4$.

If we take as our unperturbed setup a homogeneous neutrino-antineutrino gas with a single flavor, then $\rho_{12}=\langle a^\dagger_2 a_1\rangle=\langle a^\dagger_\bp a_\bp\rangle=\rho_\bp$ and similarly for $\overline\rho$, so the expressions simplify considerably to
\begin{equation}
    h_\bp=-\overline{h}_\bp=\sqrt{2}\GF\sum_{\bp'}(\rho_{\bp'}-\overline\rho_{\bp'})(T_{\bp\bp,\bp'\bp'}-T_{\bp\bp',\bp'\bp}).
\end{equation}
A perturbation need not be homogeneous. A pairing instability may well couple particles with momenta which are not opposite; this happens certainly in systems that break time reversal invariance~\cite{Fulde:1964zz, Larkin:1964wok}. 

While this is not the case here, we will nonetheless consider for generality the evolution of an inhomogeneous order parameter
\begin{equation}
\Delta_{\bp,\bq}=\bigl\langle a_{\bp+\frac{\bq}{2}}b_{-\bp+\frac{\bq}{2}}\bigr\rangle,
\end{equation}
and only later specialize to a homogeneous one. The matrix $\Phi$ then becomes
\begin{equation}
    \Phi_{\bp,\bq}= \sum_{\bp'}\Delta_{\bp',\bq} V_{\bp,\bp';\bq},
\end{equation}
where 
\begin{equation}\label{eq:scatt-amp}
    V_{\bp,\bp';\bq}=\sqrt{2}\GF \bigl(T_{\bp+\frac{\bq}{2},-\bp+\frac{\bq}{2};-\bp'+\frac{\bq}{2},\bp'+\frac{\bq}{2}}-T_{\bp+\frac{\bq}{2},\bp'+\frac{\bq}{2};-\bp'+\frac{\bq}{2},-\bp+\frac{\bq}{2}}\bigr)
\end{equation}
is the scattering amplitude for the neutrino and antineutrino participating in the pairing.
Therefore, the instability equation for $\Delta_{\bp,\bq}$ has the form
\begin{equation}\label{eq:dispersion_relation}
    i \dot{\Delta}_{\bp,\bq}=\bigl(\epsilon_{\bp+\frac{\bq}{2}}+\overline{\epsilon}_{-\bp+\frac{\bq}{2}}\bigr)\Delta_{\bp,\bq}
    +\bigl(1-\rho_{\bp+\frac{\bq}{2}}-\overline\rho_{-\bp+\frac{\bq}{2}}\bigr)
    \sum_{\bp'}\Delta_{\bp',\bq}V_{\bp,\bp';\bq},
\end{equation}
where we denote by $\epsilon=E+h$ and $\overline\epsilon=E-h$ the energies for neutrinos and antineutrinos already renormalized. A formal derivation of this equation in the neutrino context was first given in Ref.~\cite{Volpe:2013uxl}.

Notice that the scattering amplitude in Eq.~\eqref{eq:scatt-amp} is actually ill-defined because the theory of a contact interaction is not renormalizable. Therefore, we should introduce a momentum cutoff $\Lambda$. If we denote by $\widetilde{\sum}$ the sum extended only to momenta $|\bp|<\Lambda$, and by $\widetilde{V}_{\bp,\bp';\bq}(\Lambda)$ the bare tree-level scattering amplitude appearing in the Hamiltonian, then the correct instability equation is
\begin{equation}\label{eq:dispersion_relation_renormalized}
    i\dot{\Delta}_{\bp,\bq}=\bigl(\epsilon_{\bp+\frac{\bq}{2}}+\overline\epsilon_{-\bp+\frac{\bq}{2}}\bigr)\Delta_{\bp,\bq}+\bigl(1-\rho_{\bp+\frac{\bq}{2}}-\overline\rho_{-\bp+\frac{\bq}{2}}\bigr)\mathop{\widetilde\sum}\limits_{\bp'}\Delta_{\bp',\bq}\widetilde{V}_{\bp,\bp';\bq}(\Lambda).
\end{equation}
The tree-level scattering amplitude should be defined in such a way that the physical scattering amplitude in vacuum is equal to $V_{\bp,\bp';\bq}$. The reason this procedure is important is that the sum over $\bp'$ is otherwise formally divergent, since it receives contributions from states with arbitrarily large momenta. To second order in $\GF$, the bare amplitude is related to the physical one as
\begin{equation}
    \widetilde{V}_{\bp,\bp';\bq}=V_{\bp,\bp';\bq}-\mathop{\widetilde\sum}\limits_\bk \frac{V_{\bp,\bk;\bq} V_{\bk,\bp';\bq}}{E_{\bp+\frac{\bq}{2}}+E_{-\bp+\frac{\bq}{2}}-E_{\bk+\frac{\bq}{2}}-E_{-\bk+\frac{\bq}{2}}}.
\end{equation}
When the second term is consistently included, the divergence at large momenta in Eq.~\eqref{eq:dispersion_relation_renormalized} is removed~\cite{Gorkov:1061}. In practice, we may use the original Eq.~\eqref{eq:dispersion_relation} and cut it off at an energy scale comparable with the typical energy of the neutrinos in the medium. The dependence on the cutoff is only logarithmic and does not affect our conclusions.

\section{Dispersion relation for superfluid instability}

\subsection{Instability condition}

We may now derive a dispersion relation for an unstable mode with the usual methods. We seek a solution of Eq.~\eqref{eq:dispersion_relation} in the form $\Delta_{\bp,\bq}\propto e^{-i\Omega t}$ and obtain
\begin{equation}\label{eq:order_parameter}
    \Delta_{\bp,\bq}=\frac{1-\rho_{\bp+\frac{\bq}{2}}-\overline\rho_{-\bp+\frac{\bq}{2}}}{\Omega-\epsilon_{\bp+\frac{\bq}{2}}-\overline\epsilon_{-\bp+\frac{\bq}{2}}}\sum_{\bp'}\Delta_{\bp',\bq}V_{\bp,\bp';\bq}.
\end{equation}
Our physical argument against an instability can be now formulated more rigorously. The order of magnitude of the left-hand side is $\Delta$, the amplitude of the unstable eigenmode, whereas on the right-hand side, the denominator is large, with an order of magnitude $E_\bp$, the neutrino kinetic energy. Therefore, the right-hand side is of order $\Delta \mu/E_\bp$, where $\mu=\sqrt{2}\GF n_\nu$ is the scale for the neutrino-neutrino refractive energy, i.e.,  the right-hand side is smaller by a factor $\mu/E_\bp$. In a typical SN environment with $\mu\sim 1\,\mathrm{cm}^{-1}\sim 10^{-4}\,\mathrm{eV}$ and $E_\bp\sim 10\,\mathrm{MeV}$, this factor is $\mu/E_\bp\sim 10^{-11}$. Therefore, the two sides cannot be matched by many orders of magnitude. 

This mismatch of scales can be overcome only by a singularity in the integrand on the right-hand side. We use a specific example to illuminate how such a singularity may arise. We assume homogeneity ($\bq=0$) and we will also neglect in the denominator the small correction $h_\bp$ relative to the much larger kinetic energy $E_\bp$. The order parameter thus becomes
\begin{equation}\label{eq:consistency_Delta}
    \Delta_\bp=\frac{1-\rho_\bp-\overline{\rho}_{-\bp}}{\Omega-2E_\bp}\sum_{\bp'}\Delta_{\bp'}V_{\bp,\bp'},
\end{equation}
where we have dropped the suffix $\bq$ which must be evaluated at $\bq=0$. With $K_\bp=\sum_{\bp'} \Delta_{\bp'} V_{\bp,\bp'}$, the consistency condition becomes
\begin{equation}\label{eq:consistency_condition}
    K_\bp=\sum_{\bp'} V_{\bp,\bp'} K_{\bp'}\frac{1-\rho_{\bp'}-\overline\rho_{-\bp'}}{\Omega-2E_{\bp'}}.
\end{equation}
For a generic complex $\Omega=\Omega_R+i\gamma$, the integrand on the right-hand side has no divergence. We only consider the case of a positive $\gamma$; for negative $\gamma$, the integrals must be defined with a special procedure of analytical continuation~\cite{Fiorillo:2023mze,Fiorillo:2024bzm,Fiorillo:2025zio} that yields Landau-damped modes, which we are here not interested in.

In the limit $\gamma\to 0$, the dominant divergence is at $2E_{\bp'}=\Omega_R$. As explained earlier, the apparent divergence of the integral at large $|\bp'|$ is removed by a cutoff-limited integration and a renormalized scattering amplitude. In practice, we may thus interpret the integration over $\bp'$ as limited by a cutoff $\Lambda$, which only enters logarithmically
\begin{equation}
    \sum_{\bp}=\int \frac{d\bn\, p^2 dp}{(2\pi)^3}\simeq  \frac{ \Omega_R^2}{32\pi^3}\int_{\frac{\Omega_R}{2}-\Lambda}^{\frac{\Omega_R}{2}+\Lambda} dE_{\bp}\int d\bn,
\end{equation}
where we evaluate the prefactor $p^2$ at the position of the singularity $p\simeq \Omega_R/2$. We denote by $d\bn$ the integral over direction $\bn=\bp/p$ of momentum. The cutoff must be chosen of the same order as the kinetic energy, $\Lambda\sim \Omega_R$.

Thus, for $\gamma\to 0$, the integrand has a pole close to the real axis $2E_{\bp'}=\Omega_R+i\gamma$. On the other hand, this is generally not enough to cause the integral itself to become large. If the function $V_{\bp,\bp'} K_{\bp'}(1-\rho_{\bp'}-\overline\rho_{-\bp'})$ is regular across the pole, then the integral over energy may be evaluated near the pole itself, at $E_{\bp'}=\Omega_R/2$, so that
\begin{equation}
    \int_{\frac{\Omega_R}{2}-\Lambda}^{\frac{\Omega_R}{2}+\Lambda}\frac{dE_{\bp'}}{\Omega_R-2E_{\bp'}+i\gamma}=-\frac{i\pi}{2}.
\end{equation}
The real part of this integral vanishes by symmetry, while the imaginary part remains limited and has no significant enhancement for $\gamma\to 0$. Therefore, the original mismatch of scales between left and right-hand sides of Eq.~\eqref{eq:order_parameter} can never be removed.

The only way around this argument is for the singularity in the denominator $\Omega_R=2E_{\bp'}$ to coincide with a discontinuity or singularity in the numerator. This is the mathematical formulation of
the physical argument presented in Sec.~\ref{sec:superconductivity}. If the function $1-\rho_{\bp'}-\overline{\rho}_{-\bp'}=f(E_\bp,\bn)$ is discontinuous at $E_\bp=\Omega_R/2$ for some direction, then along that direction we must perform the integral over energy as
\begin{equation}
    \int_{\frac{\Omega_R}{2}-\Lambda}^{\frac{\Omega_R}{2}+\Lambda}dE_{\bp'}\frac{f(E_{\bp'},\bn')}{\Omega_R-2 E_{\bp'}+i\gamma}\simeq \log\left(\frac{2\Lambda}{\gamma}\right)\frac{f_-(\bn') - f_+(\bn')}{2}-\frac{i\pi[f_{-}(\bn')+f_+(\bn')]}{4},
\end{equation}
where $f_\pm(\bn')=\lim_{\epsilon\to 0}f(\Omega_R/2 \pm \epsilon,\bn')$.
For $\gamma \to 0$, the real part of the integral diverges, so despite the feebleness of the interaction, one can always find a sufficiently small value of $\gamma$ satisfying the equality Eq.~\eqref{eq:consistency_condition}. Notice also that if the discontinuity is that of a sign function, $f_-(\bn')=-f_+(\bn')$, the imaginary part of the integral naturally vanishes. This is exactly what happens in the standard BCS instability, where $f(E_\bp,\bn)=1-\rho_{\bp}-\rho_{-\bp}$, and at the Fermi surface, this function jumps discontinuously from $-1$ to $+1$, ensuring a negligible imaginary part and the existence of an instability.

\subsection{Spurious instabilities from discretized energies}

We finally turn to the connection between these results and the findings of Ref.~\cite{Huang:2026fcb}. The main difference, that we believe explains their detection of instabilities, is their discretization of energy. A discrete set of energies furnishes a singular density of states at each energy level, which can then be regarded as a singular Fermi surface, allowing many states to participate in separate instabilities.

To illuminate this interpretation, we examine a simplified version of our dispersion relation in Eq.~\eqref{eq:consistency_Delta} that exhibits this feature. We use an extreme case of directional discretization in the form of only two beams $\bn_1$ and $\bn_2$. We also take a very simple choice for the scattering amplitude $V_{E\bn_i;E' \bn_j}=0$;
introducing an on-diagonal scattering amplitude would only lower the growth rate, making our point less transparent. With this simple choice, the interaction mixes the two directions maximally. Modeling the structure of Eq.~\eqref{eq:consistency_Delta} schematically, we therefore use the equations
\begin{subequations}\label{eq:continuous_equations}
\begin{eqnarray}
    \Delta_{E,\bn_1}&=&\mu\,\frac{f_1(E)}{\Omega-2E}\int dE' \Delta_{E',\bn_2}, \\
    \Delta_{E,\bn_2}&=&\mu\,\frac{f_2(E)}{\Omega-2E}\int dE' \Delta_{E',\bn_2},
\end{eqnarray}
\end{subequations}
where $f_i(E)=1-\rho_{E ,\bn_1}-\overline\rho_{-E,\bn_2}$. 
The scattering amplitude has been taken energy-independent and encapsulated in the coupling constant $\mu\sim \sqrt{2}\GF n_\nu$. The simplified nature of these assumptions is irrelevant to our conclusions, namely the emergence of spurious instabilities when a discrete set of energy levels is assumed. In fact, the structure of these equations directly coincides with the one studied in Ref.~\cite{Huang:2026fcb}.

If we integrate both equations over energy, denoting $\Delta_i=\int dE \Delta_{E,\bn_i}$, we recover a self-consistency condition
\begin{equation}
    \Delta_1= \mu \Delta_2\int dE\frac{f_1(E)}{\Omega-2E}=\mu^2 \Delta_1 \int dE \frac{f_1(E)}{\Omega-2E}\int dE' \frac{f_2(E')}{\Omega-2E'},
\end{equation}
leading to the dispersion relation
\begin{equation}
    \int dE \frac{f_1(E)}{\Omega-2E} \int dE' \frac{f_2(E')}{\Omega-2E'}=\frac{1}{\mu^2}.
\end{equation}
For such a dispersion relation, all of our previous arguments apply, suggesting that a solution will not be found in the continuum limit. Indeed, the right-hand side is of the order of $\mu^{-2}$, while the left-hand side is of the order of $E^{-2}\ll \mu^{-2}$. Therefore, a solution can be found only if one of the integrals diverges, again requiring a singular surface.

As an explicit test, we examine a version of this setup that is discretized in energy so that Eq.~\eqref{eq:continuous_equations} becomes 
\begin{subequations}\label{eq:discrete_equations}
\begin{eqnarray}
    \Delta_{E_n,\bn_1}&=&\mu\,\frac{f_1(E_n)}{\Omega-2E_n}\sum_m \delta E \Delta_{E_m,\bn_2},\\
    \Delta_{E_n,\bn_2}&=&\mu\,\frac{f_2(E_n)}{\Omega-2E_n}\sum_m \delta E \Delta_{E_m,\bn_1}.
\end{eqnarray}
\end{subequations}
Notice the crucial role of the energy interval $\delta E=E_{\rm max}/N_E$,  which renormalizes the strength of the self-interaction among discrete levels and is required to attain a proper continuum limit; here $N_E$ is the number of energy bins.

\begin{figure}
    \centering
    \includegraphics[width=0.9\textwidth]{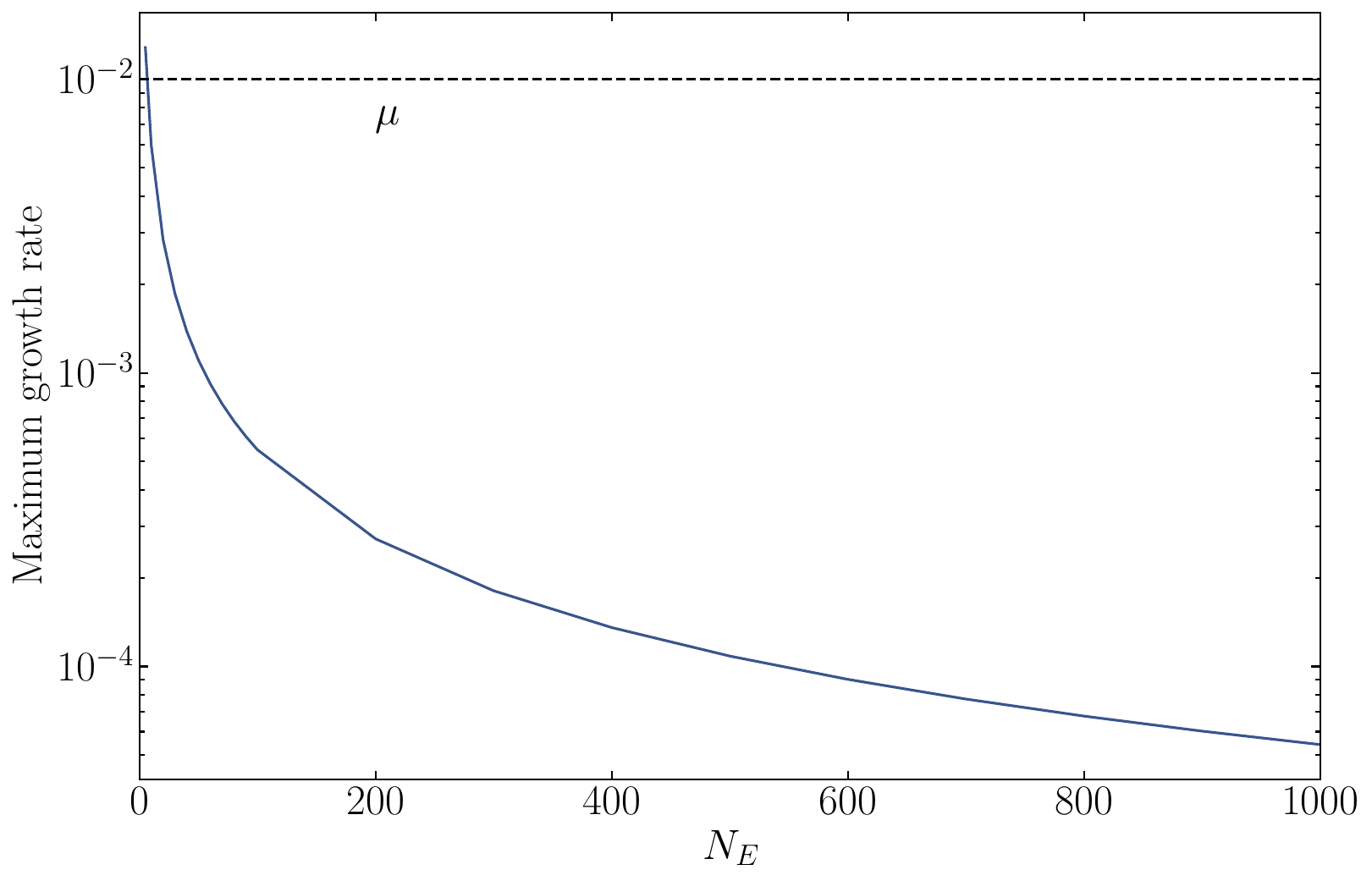}
    \caption{Maximum growth rate of the unstable eigenmodes as $N_E$ increases.}
    \label{fig:max_growth}
\end{figure}

We initialize the system with $f_1(E)=-f_2(E)=E^2 e^{-E}$. We use a dimensionless variable for $E$, with $E_{\rm max}=10$, and consider $\mu=10^{-2}$ to mimic the very small self-interaction strength in comparison with the neutrino kinetic energy. The system of Eqs.~\eqref{eq:discrete_equations} can then be solved numerically to find the eigenmodes. As in Ref.~\cite{Huang:2026fcb}, we find a host of unstable modes, but crucially, their growth rates generally decrease as the number of bins, $N_E$, increases. 

This effect is seen explicitly in Fig.~\ref{fig:max_growth}, where we show only the maximum growth rate, and we compare it with $\mu$,
the strength of the self-interaction. As $N_E$ increases, the growth rate decreases as $N_E^{-1}$, an effect coming entirely from $\delta E$, which renormalizes the strength of the self-interaction by a factor $N_E^{-1}$. Therefore, the instability disappears as $\delta E\to 0$. 

The authors of Ref.~\cite{Huang:2026fcb} also mention the possibility of persisting instabilities found when a single energy bin is taken to have a negative $f_1(E)$. This behavior probably corresponds precisely to our argument: as the number of bins increases, taking a negative value only in one of them, introduces a discontinuity in the occupation numbers, causing precisely the kind of singularity that can trigger an instability for an arbitrarily weak interaction.

\section{Discussion}

The study of collective neutrino flavor evolution has shown repeatedly that overlooked instabilities can have significant consequences on the neutrino plasma itself and its environment. Therefore, the recent proposal of a neutrino--antineutrino pairing instability~\cite{Huang:2026fcb} has to be examined in detail. In this work, we have provided a comprehensive description of such pairing phenomena from the natural perspective of superfluid instabilities.

Our main argument against the appearance of pairing instabilities in realistic astrophysical environments relies on a comparison of energy scales. The formation of a bound state from neutrinos with large kinetic energy requires a comparable potential energy, to ensure that a superposition of plane waves can form a coherent eigenstate. As a main exception, we have highlighted the case with a large reservoir of equal-energy states, which however requires a sharp Fermi surface. For neutrino--antineutrino pairing, this cannot happen even at the conceptual level in thermal equilibrium, since, if neutrinos are degenerate, antineutrinos are suppressed and have no Fermi surface. In the practical astrophysical settings of supernovae and neutron-star mergers, neutrino temperatures far exceed the self-interaction strength, and so pairing cannot happen in any case.

However, at the level of toy models, pairing instabilities can certainly occur. As shown in Ref.~\cite{Huang:2026fcb}, the resulting nonlinear behavior is conditionally periodic if a small, discrete number of energy levels and directions is selected, and becomes ever more complex as this number increases. For two directions only, their system shows pendulum-like behavior, similar to other systems of few degrees of freedom, either by construction \cite{Samuel:1993uw, Hannestad:2006nj, Padilla-Gay:2021haz, Liu:2025muc, Fiorillo:2026ybk}, or caused by conservation laws~\cite{Raffelt:2011yb, Pehlivan:2011hp, Johns:2019izj, Padilla-Gay:2022wck, Johns:2023xae, Fiorillo:2023mze, Fiorillo:2023hlk, Fiorillo:2023ajs}, or by self-selection from a resonant mechanism for weak instabilities~\cite{Fiorillo:2026vfo}. These possibilities were systematically explored in Ref.~\cite{Fiorillo:2026lyz}. 

Specifically for pairing instabilities, however, a fundamental obstacle is that a discrete set of modes can never be selected. In the dispersion relation, unoccupied modes contribute with the opposite sign as occupied ones. Therefore, one can never select a finite set of modes to participate in the dispersion relation and in the eigenmodes---the continuum nature of energy and direction is here fundamental. In particular, a discrete set of energies directly mimics a series of Fermi surfaces, therefore showing a full array of instabilities, which however are of a spurious~nature. It thus appears that the possibility of a pairing instability can after all be neglected in realistic environments. However, our argument is not as simple as one might have naively expected. The original reasoning e.g.~in Ref.~\cite{Sigl:1993ctk} that the pairing function oscillates rapidly is not sufficient, as evidenced by the very existence of superconductivity. 

Our present investigation somewhat continues our series of studies~\cite{Fiorillo:2024wej, Fiorillo:2024fnl} to examine the validity of several assumptions in the derivation of the original neutrino quantum kinetic equation~\cite{Sigl:1993ctk}. In Ref.~\cite{Fiorillo:2024fnl}, we showed that the gradients in the neutrino potential naturally produced in flavor instabilities, and neglected in the original derivation~\cite{Sigl:1993ctk}, do affect the refractive neutrino energy, but ultimately their impact on the neutrino kinetic energy can be neglected. (This phenomenon directly corresponds to the change in neutrino kinetic energy by flavomon emission~\cite{Fiorillo:2025npi}.) In Ref.~\cite{Fiorillo:2024wej}, we showed that collective neutrino--antineutrino conversions, corresponding to a particle--antihole condensate $\langle a^\dagger_{\beta,\bp} b_{\alpha,\bp}\rangle$, cannot occur due to helicity conservation, justifying the original neglect of this condensate. Finally, in the present work, we have shown that also particle--particle and particle--antiparticle condensates can be neglected, due to the lack of sharp and coincident Fermi surfaces in the neutrino and antineutrino distributions.

\section*{Acknowledgments}

The work of DFGF is funded by the European Union (ERC, NEPTUNE, No.\ 101301124). Views and opinions expressed are those of the author(s) only and do not necessarily reflect those of the European Union or the European Research Council. Neither the European Union nor the granting authority can be held responsible for them. DFGF thanks the organizers and participants of the workshop ``Collective Neutrino Oscillations in Supernovae and Neutron Star Mergers,'' and especially Shih-Jie Huang and Meng-Ru Wu, for fruitful conversations and correspondence.
DFGF acknowledges support by Istituto Nazionale di Fisica Nucleare (INFN) through the Theoretical Astroparticle Physics (TAsP) project. GGR acknowledges support by the German Research Foundation (DFG) through the Collaborative Research Centre ``Neutrinos and Dark Matter in Astro- and Particle Physics (NDM),'' Grant SFB--1258--283604770, and under Germany’s Excellence Strategy through the Cluster of Excellence ORIGINS EXC--2094--390783311. GS acknowledges support by the 
DFG under Germany’s Excellence Strategy---EXC 2121
``Quantum Universe''---390833306.

\bibliographystyle{JHEP}
\bibliography{References}

\end{document}